# Stairway to heaven: designing for an embodied experience with satellite data

Hansen, Thea Overby[a]; Jensen, Maria Leis[a]; Tonnesen, Line Schack[a]; Løvlie, Anders Sundnes*[a]

[a] IT University of Copenhagen, Denmark
*asun@itu.dk

This paper explores the design of an interactive installation in a science center which facilitates an embodied experience with satellite data. This addresses a central concern for experience design in museums, which is the question of how to integrate technologies well in the visitor experience, sometimes referred to as "experience blend". We present the design and evaluation of a visualization triggered by movement in a physical staircase to let visitors explore data about satellites at different orbits. The evaluation demonstrates strong experience blend, and points towards similar design opportunities for other institutions interested in finding new uses for mundane pathways through their buildings.

***Keywords: embodied interaction; experience blend; museums; satellites***

## 1   Introduction

This paper explores the design of an interactive installation in a science center which facilitates an embodied experience with satellite data. Scientific data visualisations have been the topic of much research in fields such as science communication and design; and are of great importance for science centers and museums. Many data visualizations in natural science museums are centered around screen-based technologies (Hohl, 2011; J. Ma et al., 2012; K.-L. Ma et al., 2012; Schuman et al., 2022). The visitors, however, can have a hard time interpreting such visualizations (Börner et al., 2016); and museum scholars and stakeholders are concerned that screen-based technologies may have a detrimental effect on the visitor experience, distracting visitors from taking in the exhibitions (Woodruff et al., 2001; Hsi, 2003; Lyons, 2009; Petrelli et al., 2013; Løvlie et al., 2021). In Human-Computer Interaction (HCI) this has been discussed through the concept of "experience blend" (van Turnhout et al., 2012; Lange et al., 2019), which describes how mixed reality experiences can be tailored to integrate well with the existing museum environment, including the visitors' expectations towards behavior in the museum and the need to support the museum's communication goals. In this study we turn to embodied interaction, a field of longstanding interest in both HCI (Dourish, 2001; Höök, 2018) and museum experience design (Levent & Pascual-Leone, 2014; Tzortzi, 2014; Hornecker

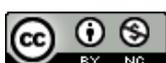

& Ciolfi, 2019; Løvlie et al., 2019), and explore the challenges involved in facilitating an embodied experience with data about satellites - which are technologies of great importance for our everyday lives, but are out of sight and far removed from our everyday lifeworlds.

Our study has been carried out in collaboration with the Copenhagen Planetarium, a science center working to convey knowledge of astronomy and space exploration. The exhibitions consist of a variety of educational as well as artistic interactive installations which explore subjects such as the Big Bang, black holes, supernovas and the moon landing. Our study was inspired by a student CubeSat satellite program and aimed to explore the use of satellite data, a topic which currently does not have much visibility in the center's exhibitions. Satellites are an important part of human presence in space, and the amount of satellites orbiting earth are increasing rapidly. However, in spite of their large quantities and their importance for everyday applications such as weather forecasts, television, Internet and navigation, it can be challenging to fully grasp the nature and extent of the network of satellites around earth. The physical properties of satellite orbits - altitude, velocity, weightlessness - are at a scale far away from the embodied experiences of human beings. For these reasons facilitating embodied experiences with satellite data is a difficult challenge, but also one with an interesting educational potential: Helping people grasp the magnitude and importance of a technological infrastructure that underlies much of our everyday communications, but which remains difficult to grasp with our everyday common sense.

Thus, this paper explores the following research question: How can we design for an embodied experience of satellite data in the Planetarium, while also designing a strong experience blend?

While this study deals with a science center, it has strong relevance for the design of experiences in other types of museums and heritage sites, because the nature of the challenges is widespread: Museums often need to present matters that are far removed from the embodied lifeworlds of their audiences, and often struggle to craft strong experience blend. As such, the insights gained from this study may be used to inform the design of experiences in a broad range of museum and heritage contexts.

## 2   Related work

The use of technology to augment visitor experiences in museums has long been an interest for design research and HCI (Hornecker & Ciolfi, 2019; Waern & Løvlie, 2022). The growing interest in embodied interaction (Dourish, 2001; Höök, 2018) is paralleled by developments in museology which put heightened emphasis on the experiences of the visitors (Vergo, 1989; Falk & Dierking, 1992; Shelton, 2013), and which have led to an increasing interested in embodied experiences in museums (Joy & Sherry, 2003; Levent & Pascual-Leone, 2014; Tzortzi, 2014; Witcomb, 2015; Steier et al., 2015; Alexander et al., 2017). Howes (2014) proposes a "sensory museology" which explores how museum experiences can engage with all the human senses. In a similar vein, multisensory experiences have been employed to explore the role of memory and emotions (Andermann & Simine, 2012).

Among museum scholars and museum stakeholders, the increasing use of technology in museums has led to concerns about technology use leading to "heads-down" experiences, in which visitors walk through the museum with their gaze and attention directed towards their mobile devices, oblivious to the exhibitions (Woodruff et al., 2001; Hsi, 2003; Lyons, 2009; Petrelli et al., 2013). Such concerns resonate with broader societal discourses on smartphone overuse and "digital detox" (Turkle, 2011; Silas et al., 2016; Panova & Carbonell, 2018; Syvertsen & Enli, 2020). Responding to such challenges, HCI scholars have explored the concept of "experience blend" (Benyon et al., 2012; van Turnhout et al., 2012; Lange et al., 2019; O'Keefe et al., 2021) discussing how mixed reality experiences can be tailored to integrate well with the existing museum environment, as well as expectations towards behavior in the museum and the need to support the museum's communication goals. Van Turnhout and colleagues (2012) initially described experience blend as an ideal for designing new experiences that seamlessly blend in with, and enhance, existing experiences. Building on this idea Lange and colleagues (2019) proposed a framework consisting of four experience design layers, representing four dimensions in which designers should consider the blend between existing practices and the new design: motivation, location, narrative, and action.

Our project has taken inspiration from Kortbek and Grønbæk (2008), who explored the challenge of integrating technology in an art museum without disrupting the artwork and visitor's experience of it. They argued for using bodily movements as the main mode of interaction rather than screen-based interfaces, in order to prevent the visitors' attention being distracted away from the art. A similar approach was presented by Hornecker & Buur (2006) describing an installation called "Clavier", in which the body's physical movements were used as an interaction device to interact with a "walkable keyboard and audio installation" located on a path in a garden. Hornecker & Buur labelled this interaction form as "spatial interaction", which they describe as an interaction where movement in physical space creates a reaction.

## 3 Method

This study has been conducted using a Research through Design methodology (Gaver, 2012; Zimmerman et al., 2007), in which insights emerge through the process of designing and evaluating prototypes. We worked in-the-wild (Rogers & Marshall, 2017) partnering with the Planetarium to evaluate prototypes with regular visitors in situ at the sicence center, allowing us to gather rich insights about how well the design integrated with the center and with the visitor's interests and behaviors.

The main goal of the project was to explore how a spatial interaction in the museum space could be used to enhance the experience of a data visualization about satellites. Furthermore, we were interested in seeing whether using the natural movement of the body in the museum space as a way of interacting with an installation could contribute to a strong experience blend. The study focused on communicating about the multitude of satellites and their placement in specific orbits around earth.

The design was evaluated through a Wizard of Oz test followed by a qualitative interview, in order to capture rich data about the experience both as seen through the test users' interactions as well as their thoughts and feelings immediately following the test. As suggested by Zimmerman and colleagues (2007) we aim not to establish scientific validity, but rather to examine the relevance of our design. Details about the tests are presented under evaluation (section 5, below).

## 4 Design process

During our initial exploration of the Planetarium we found an interesting design opportunity in the space surrounding the main staircase (Figure 1). The stairs are used by visitors to access the center's most visited attraction, a large movie theater in the building's dome using an advanced laser projection system (Skyskan) to show high quality film and data visualisation experiences about outer space. While this staircase is used by almost every visitor to the center the space of the staircase is not used to exhibit or communicate anything to visitors, which represents an unexplored opportunity. Furthermore, the staircase also facilitates an upward movement 12 meters up from the ground floor of the exhibitions, giving visitors an experience of ascending – before encountering outer space through films and simulations inside the Dome.

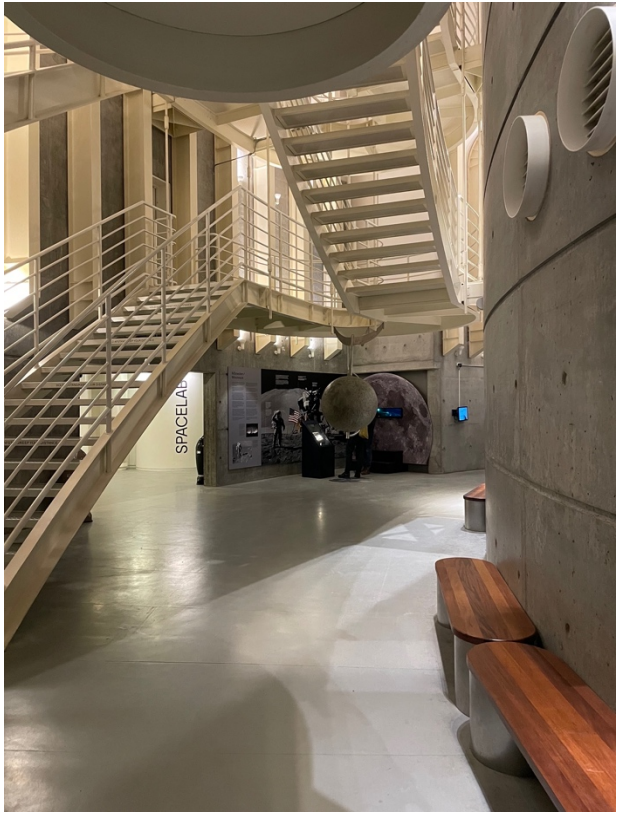

*Figure 1. The main staircase in the Planetarium in Copenhagen.*

We used rapid prototyping to create a series of lo-fi prototypes that explored different ways of interacting with satellite data, using embodied interactions, spatial interactions and gestures to activate the installation. We found that there was an interesting potential in using the visitor's physical movements to offer an embodied experience of the physical distribution of satellite orbits. We also saw a potential for using embodied interaction to facilitate exploration and play, which we included in our initial design for an installation that included miniature satellite models suspended at different levels in the stairways, which could be explored through touch to activate soundscapes (Figure 2). However, we found that this design became somewhat confusing for visitors, who did not intuitively make a connection between touching the models and the information and soundscapes accompanying them. Therefore we decided to simplify the design and integrate the use of the body more closely in the installation. This resulted in a final design where the installation was integrated into the staircase itself and visitors only had to walk the stairs without interacting with any additional interface. Thus the final design focused exclusively on using the movement up the stairs to facilitate an experience of moving out through Space.

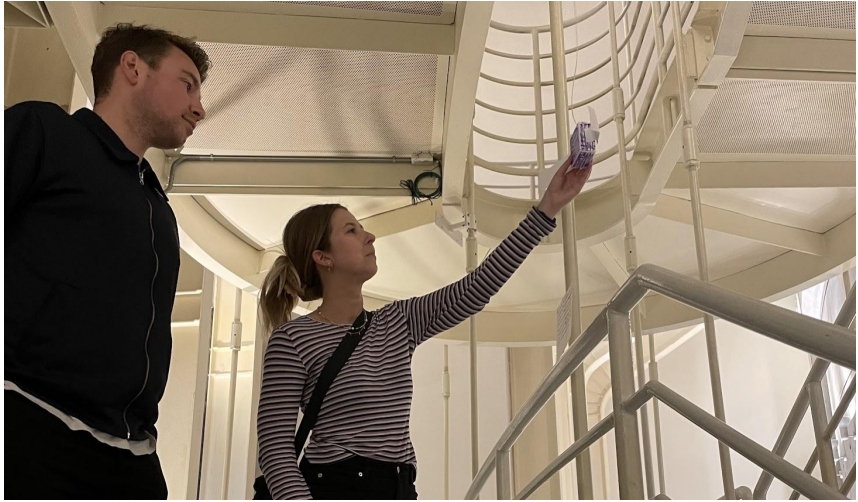

*Figure 2. Test participants interacting with a paper model of a satellite in an early prototype.*

In the final design we proposed to equip some of the steps on the staircase with sensors that activated projections on the walls as the visitors moved upwards and thus simulated moving further out in Space through different satellite orbits. These projections were accompanied by signposts on the steps indicating the altitude of the corresponding orbit (Figure 3). When no one was standing on the interactive steps the projections would show a cluster of white dots, representing the entire multitude of satellites orbiting earth. Once someone stood on one of the interactive steps, a selection of the dots would change color to represent the satellites in that specific orbit. Apart from symbolising types of satellites with different coloured dots, the projection also stated the number and type of satellites in the specific orbits. In addition, physical signposts with further information about the orbits and satellites were placed near the steps that activated the projections. The design was implemented as a lo-fi Wizard of Oz experience prototype in which the projections were triggered manually by one of the designers.

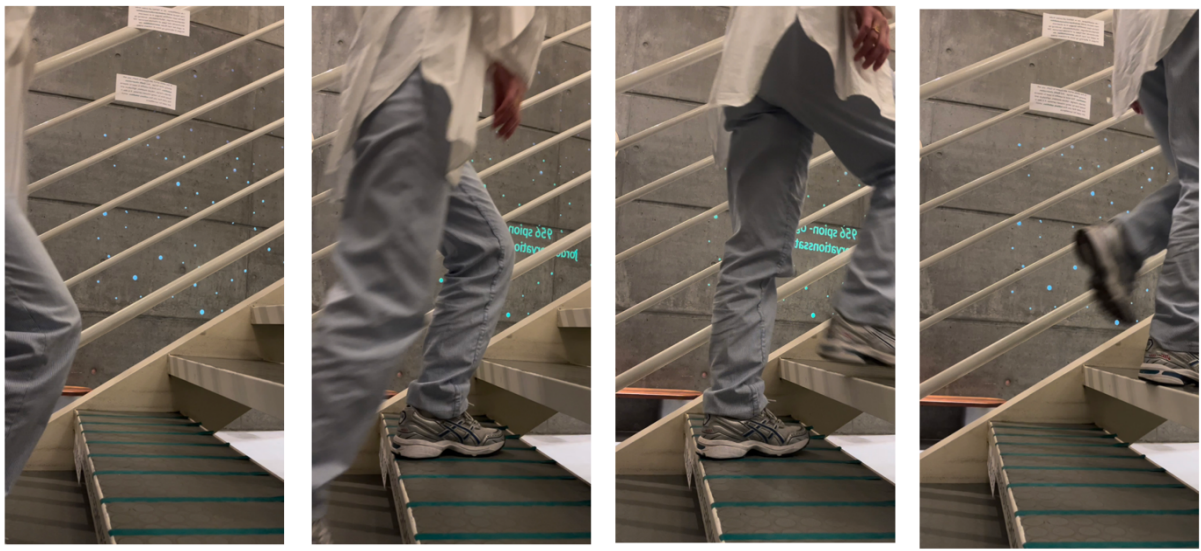

*Figure 3. Left: When the visitor is not stepping on an interactive step, the projection on the wall shows a cluster of white dots, representing the entire multitude of satellites orbiting earth. Middle: When an interactive step is stepped on the dots representing satellites in the corresponding orbit change color, and a text appears giving the number and type of satellites in this orbit. Right: When the visitor walks on, the projection reverts to its default state.*

# 5 Evaluation

We tested the design on 9 participants (3 men, 6 women, ages 18-32) on location in agreement with Planetarium. 5 of the participants were recruited beforehand and invited to visit Planetarium in order to try the prototype, while the remaining 4 were visitors which we approached on the day. We told the participants that they would be trying out an interactive design and asked them to act as they would normally do in a museum context and when interacting with exhibitions. Afterwards we interviewed each participant about their experience of the design - specifically their experience of the interaction, the movement up the stairs and the interplay between these two aspects. The interviews were recorded and transcribed and thereafter individually analyzed by the authors and relevant quotes and insights were extracted. Afterwards we compared and discussed our analysis until we reached consensus on the relevant insights. These insights were divided inductively into overarching themes. Finally, we re-visited the categories and regrouped them into themes.

In the following we will briefly reflect upon the results, highlighting two themes: The link between participants' movements and the spatial information in the prototype, and their awareness of how their movements triggered interactions with the prototype.

## 5.1 Participants' movements

When asked about their movement up the stairs, participants highlighted that the movement gave them a physical feeling of moving out through space. Nearly every participant mentioned that it contributed to a better understanding of where the different satellites are positioned compared to Earth and each other, and that it added an extra "layer" to the experience. A participant said: "I actually think that it was a nice effect to be able to move up the stairs and at the same time see the distances. You almost felt like you were moving through space". Another highlighted that the movement contributed to the overall narrative: "I just think that is a nice part of the narrative that you move upwards. It has something to do with the story, that you can see one satellite down here and the next one up there. It did a lot for the story". Participants emphasized that the movement up the stairs helped them understand how different satellites are located - and thus they experienced a close link between their movement and the narrative in the installation.

## 5.2 Awareness of interaction possibilities

Somewhat to our surprise, almost none of the participants noticed that the changes in the projections happened because they stepped on specific steps in the staircase. Even the few who did notice, only did so after having explored the prototype for a while. Some participants indicated that they thought the projections were linked to their movements in general, rather than specific steps: "It felt as if it [the projections] followed you on the way up, it worked really nicely." This suggests that our test participants found it quite intuitive and "natural" that the projections changed as they moved on the stairs. Among the few who discovered that specific steps triggered the interaction, some started to play with the installation: challenging the system's ability to react to their steps by quickly stepping on and off the interactive steps, checking to see how quickly the system would react. This seems to indicate that there is a potential for designing for playful interactions with this setup, as described by one participant: "It would also be a bit boring if it [the interaction] was too obvious. It is the aspect of having to explore on your own that makes it fun when you finally discover it".

# 6 Discussion and implications for future work

The design presented in this paper employs strategies similar to Kortbek and Grønbæk's (2008) ideas for using the body as an interaction device to redirect visitors' attention away from interfaces and towards the main content of the museum exhibitions. In our design, the use of spatial and embodied interactions led many users to accept the experience without fully understanding how their actions triggered the changes, resulting in an integrated experience that emphasized the communication of scientific data instead of having the visitor's primary focus be on the interaction with the technology. Even for those participants who failed to realize how their own movements triggered the changes in the projections, this did not seem to make the experience less enjoyable.

Table 1 summarises the experience blend along the model proposed by Lange and colleagues (2019), demonstrating a strong blend with the pre-existing experience. The prototype addresses the same motivations and narratives as the existing exhibition. The 'action' dimension points to an issue that might be worth some consideration if the prototype were to be further developed and implemented as a fully functional system, as it is not entirely certain whether this would lead visitors' behavior to change and disrupt the flow of people in the stairs. However, in the current version the projections are so simple that they afford a meaningful experience even for visitors who only glance at the projections briefly in passing, thus making it possible to take in the experience while traversing the staircase at a normal pace. The museum might also wish to give visitors the opportunity to explore more detailed information, for instance by adding interactive displays with educational material about satellites and related themes such as space junk. This might require careful design and testing in order to avoid disrupting the flow in the stairs - for instance by placing such displays at well-chosen spots at the bottom and top of the stairs, where visitors can stop without blocking the path for others.

*Table 1. Experience blend of the Stairway to Heaven prototype*

|  | Existing experience | Blend | New experience |
|---|---|---|---|
| **Motivation** | Experiencing and learning about astronomy and space exploration | Strong | Visualising and learning about satellites |
| **Location** | Empty staircase | Strong | Projections on the surrounding walls, triggered by sensors in the stairs |
| **Narrative** | Exhibition and films about space exploration, but little information about satellites | Strong | Visualisation and information about the types of satellites orbiting earth |
| **Action** | Walking up the stairs to get to the Dome | Medium | Watching the projections on the walls while ascending the stairs; possibly stopping to study them, or going back and forth to play with the sensors |

However, the model obscures what is arguably the most valuable aspect of the design: Turning a mundane staircase, which currently does not contain any exhibition material, into an aesthetic and educational experience. While this design blends well with the overall visitor experience, if we focus more narrowly on the "experience" of traversing the staircase the design does not so much blend with the current (non-)experience, as rather craft an entirely new experience into what is currently just a nondescript pathway. This offers value both for visitors and for the science center, and indicates a design strategy that might be applied in other venues: Using mundane pathways such as stairs, corridors and elevators to craft immersive visualisations that visitors can explore while traversing the physical space. It is easy to imagine similar designs for other science centers or museums: For instance using the building's mundane pathways to allow visitors to move through visualisations on varying scales, e.g. simulating a journey to the earth's interior, or through the structure of a molecule, and so on. If the building's structure contains many interconnected pathways - corridors, stairways etc - one might imagine larger installations that use multiple pathways to allow visitors to explore complex structures such as an ant hill or the inside of the human body.

Such a design strategy of "reappropriating" mundane pathways might also be applicable in other types of locations that include complex networks of pathways, such as metro stations and train stations, airport terminals etc. Such designs might serve artistic, educational as well as commercial purposes.

# 7   Conclusion

The study in this paper has demonstrated the potential for utilising mundane pathways to craft novel visitor experiences with data visualisation. The study suggests an approach to facilitate an embodied understanding of scientific phenomena and data – such as distances in space and how satellites are placed in orbit around earth. Our design work underlines that spatial interactions offer the possibility to blend in with existing experiences, transforming mundane locations into educational or aesthetic experiences. This design strategy opens up a new way of thinking about pathways - adding new life and excitement to places otherwise only used for passing through. The study also indicates some concerns that should be taken into account, regarding experience blend and making visitors aware of the possibilities for interaction. More research is needed to explore these design opportunities further, including with fully functional implementations that can be tested in a variety of situations to uncover unexpected uses and test the robustness of the interactions during crowded events and other scenarios not covered by the limited study presented in this paper.

# References

Alexander, J., Wienke, L., & Tiongson, P. (2017). *Removing the barriers of Gallery One: A new approach to integrating art, interpretation, and technology – MW17: Museums and the Web 2017*. MW17: Museums and the Web 2017, Cleveland, Ohio, USA. https://mw17.mwconf.org/paper/removing-the-barriers-of-gallery-one-a-new-approach-to-integrating-art-interpretation-and-technology/index.html

Andermann, J., & Simine, S. A. (2012). Introduction: Memory, community and the new museum. *Theory, Culture & Society*, *29*(1), 3–13. https://doi.org/10.1177/0263276411423041

Benyon, D., Mival, O., & Ayan, S. (2012). Designing blended spaces. *The 26th BCS Conference on Human Computer Interaction 26*, 398–403.

Börner, K., Maltese, A., Balliet, R. N., & Heimlich, J. (2016). Investigating aspects of data visualization literacy using 20 information visualizations and 273 science museum visitors. *Information Visualization*, *15*(3), 198–213. https://doi.org/10.1177/1473871615594652

Dourish, P. (2001). *Where the Action Is: The Foundations of Embodied Interaction*. The MIT Press.

Falk, J. H., & Dierking, L. D. (1992). *The Museum Experience*. Whaleshack Books.

Gaver, W. (2012). What Should We Expect from Research Through Design? *Proceedings of the SIGCHI Conference on Human Factors in Computing Systems*, 937–946. https://doi.org/10.1145/2207676.2208538

Hohl, M. (2011). Sensual Technologies: Embodied experience and visualisation of scientific data. *Body, Space & Technology*, *10*(1), Article 1. https://doi.org/10.16995/bst.89

Höök, K. (2018). *Designing with the Body: Somaesthetic Interaction Design*. https://doi.org/10.7551/mitpress/11481.001.0001

Hornecker, E., & Buur, J. (2006). Getting a Grip on Tangible Interaction: A Framework on Physical Space and Social Interaction. *Proceedings of the SIGCHI Conference on Human Factors in Computing Systems*, 437–446. https://doi.org/10.1145/1124772.1124838

Hornecker, E., & Ciolfi, L. (2019). Human-Computer Interactions in Museums. *Synthesis Lectures on Human-Centered Informatics*, *12*(2), i–171. https://doi.org/10.2200/S00901ED1V01Y201902HCI042

Howes, D. (2014). Introduction to sensory museology. *Senses and Society*, *9*(3), 259–267. https://doi.org/10.2752/174589314X14023847039917

Hsi, S. (2003). A Study of User Experiences Mediated by Nomadic Web Content in a Museum. *Journal of Computer Assisted Learning*, *19*(3), 308–319. https://doi.org/10.1046/j.0266-4909.2003.jca_023.x

Joy, A., & Sherry, J. F., Jr. (2003). Speaking of Art as Embodied Imagination: A Multisensory Approach to Understanding Aesthetic Experience. *Journal of Consumer Research*, *30*(2), 259–282. https://doi.org/10.1086/376802

Kortbek, K. J., & Grønbæk, K. (2008). Communicating art through interactive technology: New approaches for interaction design in art museums. *Proceedings of the 5th Nordic Conference on Human-Computer Interaction: Building Bridges*, 229–238. https://doi.org/10.1145/1463160.1463185

Lange, V., van Beuzekom, M., Hansma, M., Jeurens, J., van den Oever, W., Regterschot, M., Treffers, J., van Turnhout, K., Ibrahim Sezen, T., Iurgel, I., & Bakker, R. (2019). Blending into the White Box of the Art Museum. *Proceedings of the Halfway to the Future Symposium 2019*, 1–10. https://doi.org/10.1145/3363384.3363469

Levent, N., & Pascual-Leone, A. (2014). *The Multisensory Museum: Cross-Disciplinary Perspectives on Touch, Sound, Smell, Memory, and Space*. Rowman & Littlefield.

Løvlie, A. S., Benford, S., Spence, J., Wray, T., Mortensen, C. H., Olesen, A., Rogberg, L., Bedwell, B., Darzentas, D., & Waern, A. (2019, April). The GIFT framework: Give visitors the tools to tell their own stories. *Museums and the Web 2019*. Museums and the Web 2019, Boston, MA, USA. https://mw19.mwconf.org/paper/the-gift-framework-give-visitors-the-tools-to-tell-their-own-stories/

Løvlie, A. S., Ryding, K., Spence, J., Rajkowska, P., Waern, A., Wray, T., Benford, S., Preston, W., & Clare-Thorn, E. (2021). Playing Games with Tito: Designing Hybrid Museum Experiences for Critical Play. *J. Comput. Cult. Herit.*, *14*(2). https://doi.org/10.1145/3446620

Lyons, L. (2009). Designing Opportunistic User Interfaces to Support a Collaborative Museum Exhibit. *Proceedings of the 9th International Conference on Computer Supported Collaborative Learning*, *1*, 375–384. https://doi.org/s

Ma, J., Liao, I., Ma, K.-L., & Frazier, J. (2012). Living Liquid: Design and Evaluation of an Exploratory Visualization Tool for Museum Visitors. *IEEE Transactions on Visualization and Computer Graphics*, *18*(12), 2799–2808. https://doi.org/10.1109/TVCG.2012.244

Ma, K.-L., Liao, I., Frazier, J., Hauser, H., & Kostis, H.-N. (2012). Scientific Storytelling Using Visualization. *IEEE Computer Graphics and Applications*, *32*(1), 12–19. https://doi.org/10.1109/MCG.2012.24

O'Keefe, B., Flint, T., Mastermaker, M., Sturdee, M., & Benyon, D. (2021). Designing Blended Experiences. *Designing Interactive Systems Conference 2021*, 309–321.

Panova, T., & Carbonell, X. (2018). Is smartphone addiction really an addiction? *Journal of Behavioral Addictions*, *7*(2), 252–259. https://doi.org/10.1556/2006.7.2018.49

Petrelli, D., Ciolfi, L., van Dijk, D., Hornecker, E., Not, E., & Schmidt, A. (2013). Integrating Material and Digital: A New Way for Cultural Heritage. *Interactions*, *20*(4), 58–63. https://doi.org/10.1145/2486227.2486239

Rogers, Y., & Marshall, P. (2017). Research in the Wild. *Synthesis Lectures on Human-Centered Informatics*, *10*(3), i–97. https://doi.org/10.2200/S00764ED1V01Y201703HCI037

Schuman, C., Stofer, K. A., Anthony, L., Neff, H., Chang, P., Soni, N., Darrow, A., Luc, A., Morales, A., Alexandre, J., & Kirkland, B. (2022). Ocean Data Visualization on a Touchtable Demonstrates Group Content Learning, Science Practices Use, and Potential Embodied Cognition. *Research in Science Education*, *52*(2), 445–457. https://doi.org/10.1007/s11165-020-09951-9

Shelton, A. (2013). Critical Museology: A Manifesto. *Museum Worlds*, *1*(1), 7–23. https://doi.org/10.3167/armw.2013.010102

Silas, E., Løvlie, A. S., & Ling, R. (2016). The smartphone's role in the contemporary backpacking experience. *Networking Knowledge: Journal of the MeCCSA Postgraduate Network*, *9*(6), 40–53.

Steier, R., Pierroux, P., & Krange, I. (2015). Embodied interpretation: Gesture, social interaction, and meaning making in a national art museum. *Learning, Culture and Social Interaction*, *7*, 28–42. https://doi.org/10.1016/j.lcsi.2015.05.002

Syvertsen, T., & Enli, G. (2020). Digital detox: Media resistance and the promise of authenticity. *Convergence*, *26*(5–6), 1269–1283. https://doi.org/10.1177/1354856519847325

Turkle, S. (2011). *Alone Together: Why We Expect More from Technology and Less from Each Other*. Basic Books. https://www.basicbooks.com/titles/sherry-turkle/alone-together/9780465093656/

Tzortzi, K. (2014). Movement in museums: Mediating between museum intent and visitor experience. *Museum Management and Curatorship*, *29*(4), 327–348. https://doi.org/10.1080/09647775.2014.939844

van Turnhout, K., Leer, S., Ruis, E., Zaad, L., & Bakker, R. (2012, September 26). *UX in the Wild: On Experience Blend & Embedded Media Design*. The Web and Beyond 2012 16th Chi Nederland conference, Amsterdam, Netherlands.

Vergo, P. (1989). *The New museology*. Reaktion Books.

Waern, A., & Løvlie, A. S. (Eds.). (2022). *Hybrid Museum Experiences*. Amsterdam University Press. https://www.aup.nl/en/book/9789048552849/hybrid-museum-experiences

Witcomb, A. (2015). Cultural pedagogies in the museum: Walking, listening and feeling. In M. Watkins, G. Noble, & C. Driscoll (Eds.), *Cultural Pedagogies and Human Conduct* (pp. 158–170). Routledge.

Woodruff, A., Aoki, P., Hurst, A., & Szymanski, M. (2001). Electronic Guidebooks and Visitor Attention. In D. Bearman & F. Garzotto (Eds.), *International Cultural Heritage Informatics Meeting: Proceedings from ichim01*. Archives & Museum Informatics. http://www.archimuse.com/publishing/ichim01_vol1/woodruff.pdf

Zimmerman, J., Forlizzi, J., & Evenson, S. (2007). Research Through Design As a Method for Interaction Design Research in HCI. *Proceedings of the SIGCHI Conference on Human Factors in Computing Systems*, 493–502. https://doi.org/10.1145/1240624.1240704